# The origin of the glassy magnetic dynamics of the phase segregated state in perovskites

F. Rivadulla[1]*, M. A. López-Quintela[1], J. Rivas[2]
[1]*Physical Chemistry and* [2]*Applied Physics Departments, University of Santiago de Compostela*
*15782- Santiago de Compostela (SPAIN)*

(*Corresponding author (F.R.) e-mail: qffran@usc.es)

**Abstract**.- In this paper we address many of the fundamental open questions regarding the glassy behavior of the magnetic/electronic phase segregated state in rare earth perovskites. In particular, magnetic relaxation experiments support that the collective effects (memory, ageing, etc.) are due to interparticle interactions, rather than the double-exchange vs. superexchange competition. A careful study of the non-linear susceptibility in the critical region is performed, and the critical exponents contrasted with those of conventional spin-glasses and concentrated quenched ferrofluids. The phase segregated state constitutes a sort of self-generated assembly of magnetic particles in which magnetic interaction introduces collectivity among the clusters.

The phase segregated state (PSS) that develops in many systems close to a first-order electronic phase transition is being intensively studied, mainly because of its potential relevance to colossal magnetoresistance and high temperature superconductivity [1,2].

For the particular case of manganites, there are plenty of examples in the literature of unusual relaxation dynamics and frequency dependent phenomena in the PSS which denote a certain degree of glassines (see for example chap. 13 in [2], and references therein). Among the most representative examples of this behavior, Freitas *et al.*[3] and Levy *et al.*[4] reported ageing/rejuvenation and memory effects on the resistivity and magnetization of different manganites in the PSS. The occurrence of these time-dependent phenomena, many of them similar to those reported in classical spin glasses [5], made it usual to refer to the PSS as *cluster-glass* or *spin-glass like* phase. Moreover, there is a total consensus at the time to ascribe the origin of the spin-glass like characteristics of the PSS to the frustration introduced by the competition between FM-double exchange (DE) and AF-superexchange (SE). However, these experiments only proof the existence of a sort of collective relaxation behavior, but do not provide any definitive evidence of a true spin-glass. Instead, a detailed scaling analysis must be performed at the critical region, close to the transition temperature. So, at this moment, the fundamental question of whether the PSS of manganites can be described as a classical spin-glass, or if on the other hand, it constitutes a new class of glass, remains still open.

In this paper we address this issue by an analysis of the critical behavior of the non-linear susceptibility, $\chi_{nl}$, which is proportional to the divergent spin-glass order parameter susceptibility ($\chi_{SG} \propto \varepsilon^{-\gamma}$). The main conclusions of this work is are that the glassines of the PSS is due to interparticle interactions, and that the system in the PSS regime must be considered as an assembly of interacting magnetic particles; this interaction introduces the collectivity and glassines observed in the relaxation experiments.

For simplicity, only experiments performed on a ceramic sample of $(La_{0.25}Nd_{0.75})_{0.7}Ca_{0.3}MnO_3$ will be discussed, although similar effects where observed in other compositions in the PSS (other manganites, but also rare-earth cobaltates[6]). The proximity of this composition to the metal-insulator transition [7] makes it an optimal choice for studies of magnetic relaxation phenomena. Although M(T) rises close to T* in $(La_{0.25}Nd_{0.75})_{0.7}Ca_{0.3}MnO_3$ (see Fig. 2), it does not constitute a true FM-$T_C$, as it was recently demonstrated [8]. Instead, a PSS (ferromagnetic clusters in a paramagnetic matrix) develops below T* due to the proximity to the localized-to-itinerant crossover, which restrains the magnetic correlation length $\xi$ to the size of the FM-clusters, and prevents the $\xi \rightarrow \infty$ divergence at T*.

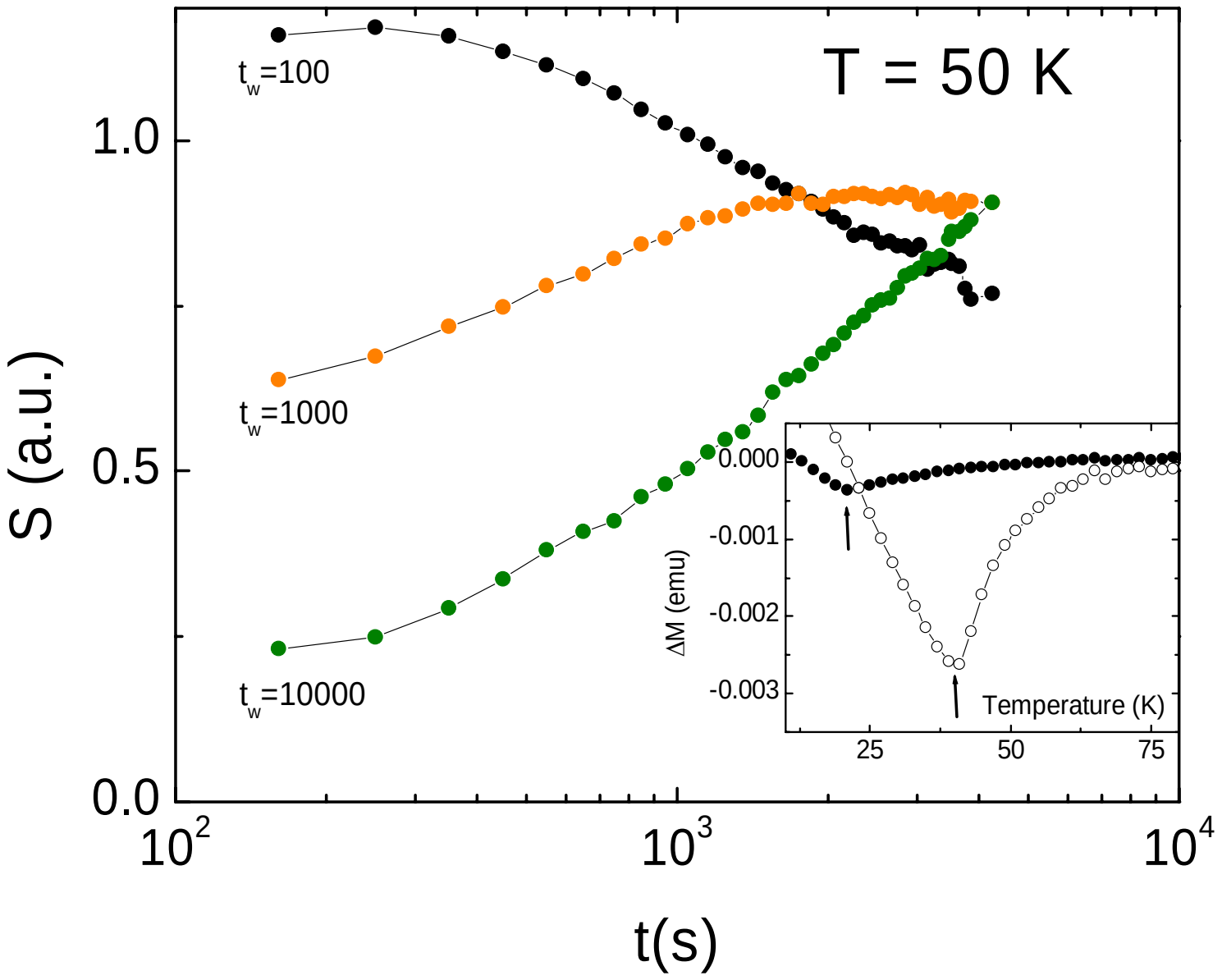

**Figure 1**: The relaxation function $S=(1/H)\{\partial M/\partial \mathrm{dln}t\}$ vs. log t at 50 K for different waiting times, $t_w$. There is a displacement of the maximum in S(t) with respect to $t_w$, due to the extra time employed by the SQUID to stabilize the temperature before starting the measurement. *Inset*: Memory effect in the dc ZFC magnetization.

The non-equilibrium dynamics and memory phenomena in the PSS are reported in figure 1. The sample was cooled at zero field down to a certain T<T*, and after a waiting time $t_w$, a field was applied and the time dependent magnetization was recorded. The function $S=(1/H)(\partial m/\partial \ln t)$ vs. log t, is plotted in figure 1 for different $t_w$. A spin-glass left unperturbed by external fields at a constant T, rearranges its spin configuration through a very slow process to reduce the domain wall energy. Within the Fisher and Huse droplet model [9] the magnetic response to an applied field after a $t_w$ is probing the polarization of domains with different characteristic length scales. A crossover at $t \approx t_w$ due to gradual excitation of larger droplets including domain walls is expected, and will show up as a maximum in S(log t) at $t \approx t_w$. As we already mentioned, similar results are available for other compositions inside the PSS region [3]. However, this behavior is not exclusive of conventional spin-glasses and has been reported in systems of concentrated magnetic particles [10], and other nanostructure magnetic materials [11], where dipolar interactions introduce a collective state and magnetic relaxation dependence like that of spin-glasses at low temperatures. So, the existence of ageing is only a proof of the existence of a collective relaxation, but not of a thermodynamic spin-glass phase.

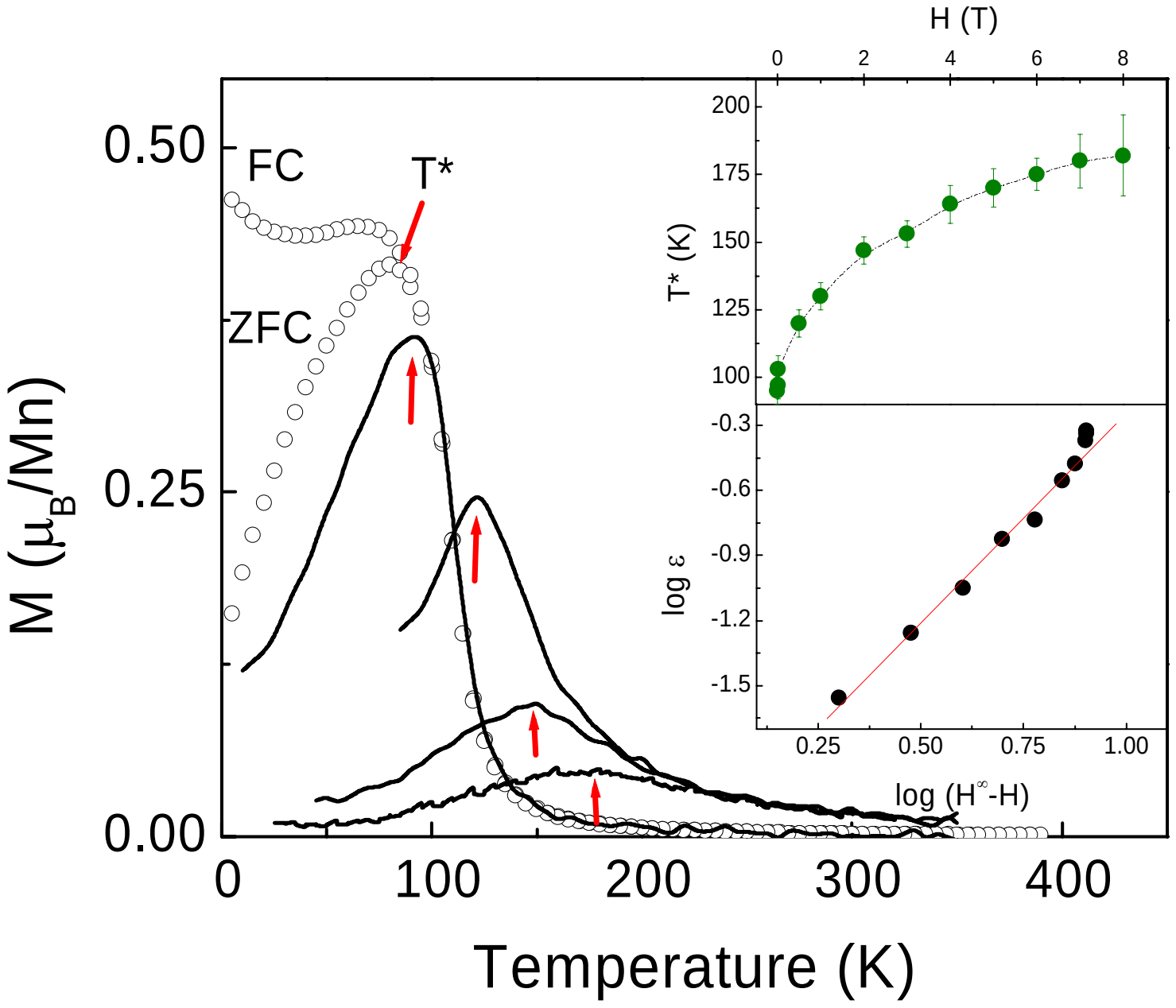

**Figure 2**: Temperature dependence of the ZFC-FC dc magnetization at 10 Oe (open symbols). The ac-ZFC curves measured with different superimposed dc fields (solid lines) are also included. *Upper inset*: The field dependence of T* extracted from the peak in the ac-ZFC susceptibility (*h*=1-10 Oe, 10 Hz) with a superimposed dc field (up to 9T). *Lower inset*: log-log plot of the reduced temperature vs. ($H^\infty$-H). The line is the linear fit from where the critical exponent ν for the correlation length was obtained.

In fact many of these effects can be perfectly qualitatively understood on the basis of a system composed by non-interacting particles with a temperature dependent distribution of relaxation times [12]. To discard this possibility and to confirm proper spin-glass dynamics, a ZFC experiments with stops during cooling at zero field must be performed. In a spin-glass or a system of interacting magnetic particles (but not in non-interacting particles), a dip appears on reheating at the temperature at which the sample was stopped under zero field. This is exactly what we observed after stopping the ZFC process at 40K and 20K (Fig.1, inset). This constitutes a very important experiment as it confirms that the memory effects observed in the PSS reflects spin-glass dynamics, whether it constitutes a classical spin-glass or the behavior is introduced by interparticle interactions. To discern between these two possibilities is the main goal of this paper, and with this aim the field dependence of the magnetization and the non-linear susceptibility at the critical region were carefully studied.

In figure 2 we show M (T,H) in ac/dc conditions. The unusually strong dependence of T* on the magnetic field is a consequence of finite size effects introduced by the limitation of $\xi$ to the size of the FM-clusters stabilized below T*, and can be described by the following size scaling law:

$$\frac{T^{\infty} - T^*}{T^{\infty}} = \varepsilon = \left(\frac{\xi}{\xi_0}\right)^{-1/\nu} \quad (1)$$

where $T^{\infty} \approx 180$ K is the $T_C$ of the infinite cluster ($\xi \rightarrow \infty$) obtained from the saturation of the T* vs. H curve (Fig. 2, upper inset). From the field dependence of $\xi$ observed in SANS experiments in the PSS [13], a phenomenological dependence $\xi \propto (H^{\infty}-H)^{-x}$ can be anticipated. In this case $H^{\infty}$ is the critical field at which $\xi \rightarrow \infty$. The critical exponent for the correlation length obtained in this way is $\nu$=1.08(2) (Fig. 1, inset), using a value of x $\approx$ 1.7, consistent with neutron scattering experiments. Numerical calculations and experimental results in fine particle systems reported a similar dependence of the $T_C$ with particle size, governed by the finite-scaling law of eq. (1) [14]. This is a strong argument in favor of a PSS composed of interacting magnetic particles whose size/interactions can be tuned by a magnetic field.

However, the definitive way to confirm the existence of a true spin-glass is to study the critical behavior of the spin-glass order parameter susceptibility and check the critical exponents through scaling. In the low field limit and in the critical region, the spin-glass order parameter susceptibility diverges of the form $\chi_{SG} \propto \varepsilon^{\gamma}$, but it has been shown to be proportional to the $H^2$ term in the static scaling expansion of the nonlinear susceptibility,

$$\chi_{nl} = \chi_0 - M/H = \chi_3 H^2 + \chi_5 H^4 + \ldots (2)$$

Geschwind, Huse, and Devlin [15] proposed an static scaling equation for the nonlinear susceptibility of spin glasses of the form

$$\chi_{nl} = H^{2\beta/(\beta+\gamma)} F\left[\frac{\varepsilon}{H^{2/(\beta+\gamma)}}\right] \quad (3)$$

This equation is based on a linear relationship between $\chi_{nl}$ and $\varepsilon$, instead of the usual logarithmic, reducing the errors in the determination of the critical exponents. A general scaling analysis according to eq. (3) was performed for isotherms in the critical region above T* (Fig. 3). Best data collapsing was obtained with $\beta$=0.30(3) and T*=97K, using $\gamma$=2.5(1) obtained from the fit of the real component of the third harmonic of the ac susceptibility (Fig.3, inset) [16].

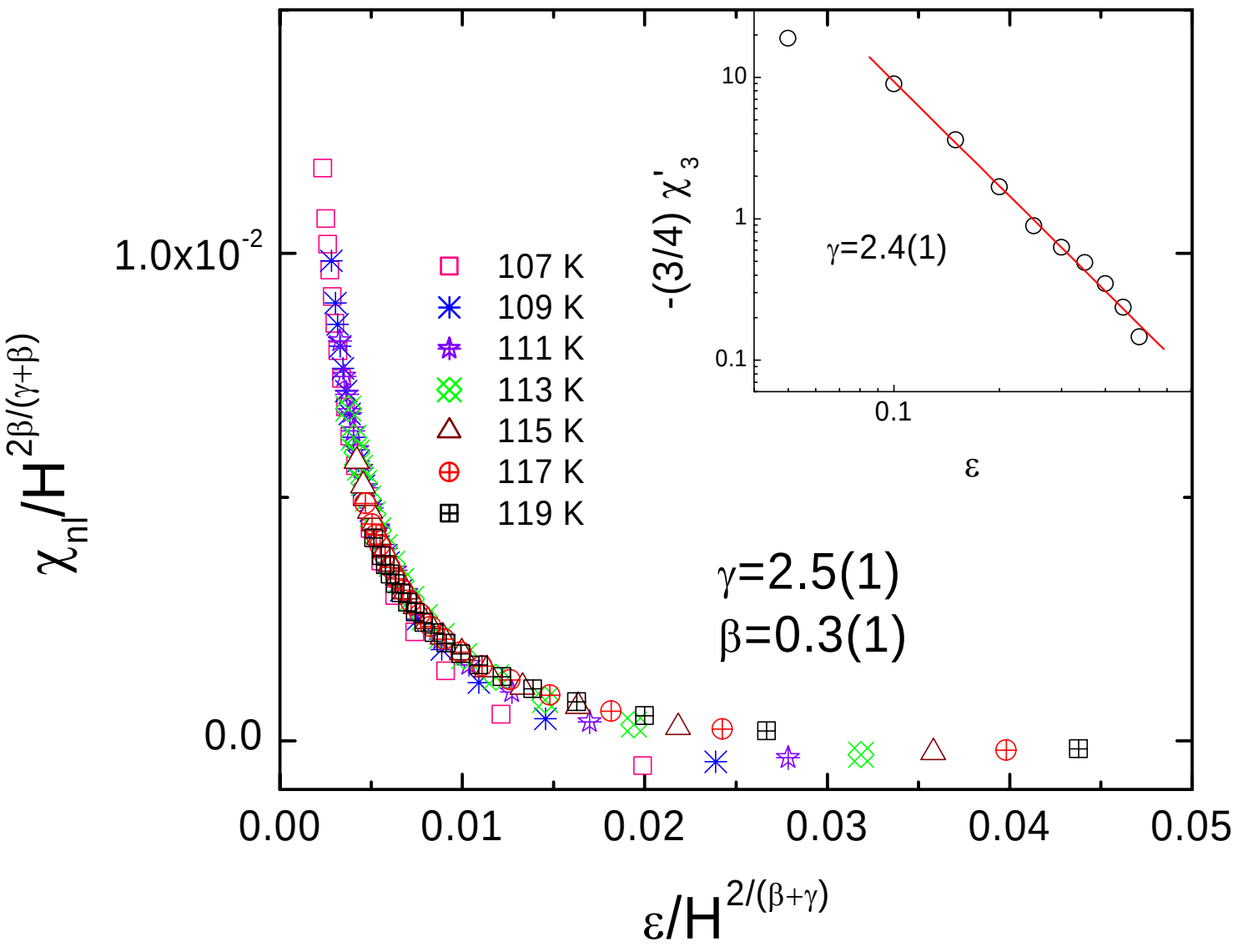

**Figure 3**: Scaling plot of the non linear susceptibility above T*. *Inset*: Critical exponent of the susceptibility derived from the non-linear component of $\chi_{ac}$ at different amplitudes.

To the best of our knowledge, this is the first time the divergent character of the nonlinear susceptibility is demonstrated in a PSS system. Applying the scaling relations between the critical exponents we got α=-1.1 and ν=1.03. The latter one agrees reasonable well with the value obtained from finite size scaling of eq. (1), supporting our analysis. On the other hand, it is difficult to ascribe the critical exponents to an universality class, and then it could be tempting to conclude that the system behaves as an interacting assembly of magnetic particles, where similar scaling collapse has been observed but with critical exponents also impossible to ascribe to an universality class [17]. However, the critical exponents we obtained experimentally are similar to these of amorphous materials classically considered as spin-glasses, like $Fe_{10}Ni_{70}P_{20}$ [18]. The problem is that there is an enormous dispersion among the values of the critical exponents reported in the literature for different spin-glasses, which makes it extremely difficult to decide to which universality class corresponds a particular spin-glass material. So in this case, scaling analysis of the critical nonlinear susceptibility is not enough to discern between a conventional spin-glass and a system of interacting magnetic particles.

On the other hand, Ulrich *et al.*[19] using Monte Carlo simulations demonstrated that in a frozen ferrofluid below its blocking temperature, dipolar interactions causes the relaxation rate, $W(t) = -(d/dt)\ln M(t)$, to decay following a power law, with an exponent *n*, which depends on the concentration and hence on the strength of the magnetic interaction. The authors interpreted the scenario described by $n \geq 1$ as a clear signature of a spin-glass phase. These theoretical predictions were recently corroborated by magnetic relaxation measurements in granular magnetic films [20]. In figure 4 we show the decay of *W(t)*, and the values of *n* obtained for various fields and temperatures, always below T*. At low temperature (≈0.2T*) and at low fields, *n* is always of the order of 1, which indicates that the relaxation of the system is governed by interactions (collective state), reminding the behavior of a true spin-glass. However, when the temperature is increased *n* is not constant, in opposition to a true spin-glass or a system of strongly interacting particles with fixed diameter and concentration. In the PSS, *n* increases continuously as T* is approached, reaching *n*=1.50(2) at 0.72T*. This experiment confirms that in the PSS an increase in the temperature towards the T* is equivalent to an increase in the concentration in a system of magnetic particles. On the other hand, the increase of the magnetization as T* is approached from below points to an increase in the number and/or the size of the particles (M is proportional to the volume and number of individual particles). So, the progressive increment of the exponent *n* reflects the increasing strength of the interparticle interactions as T* is approached.

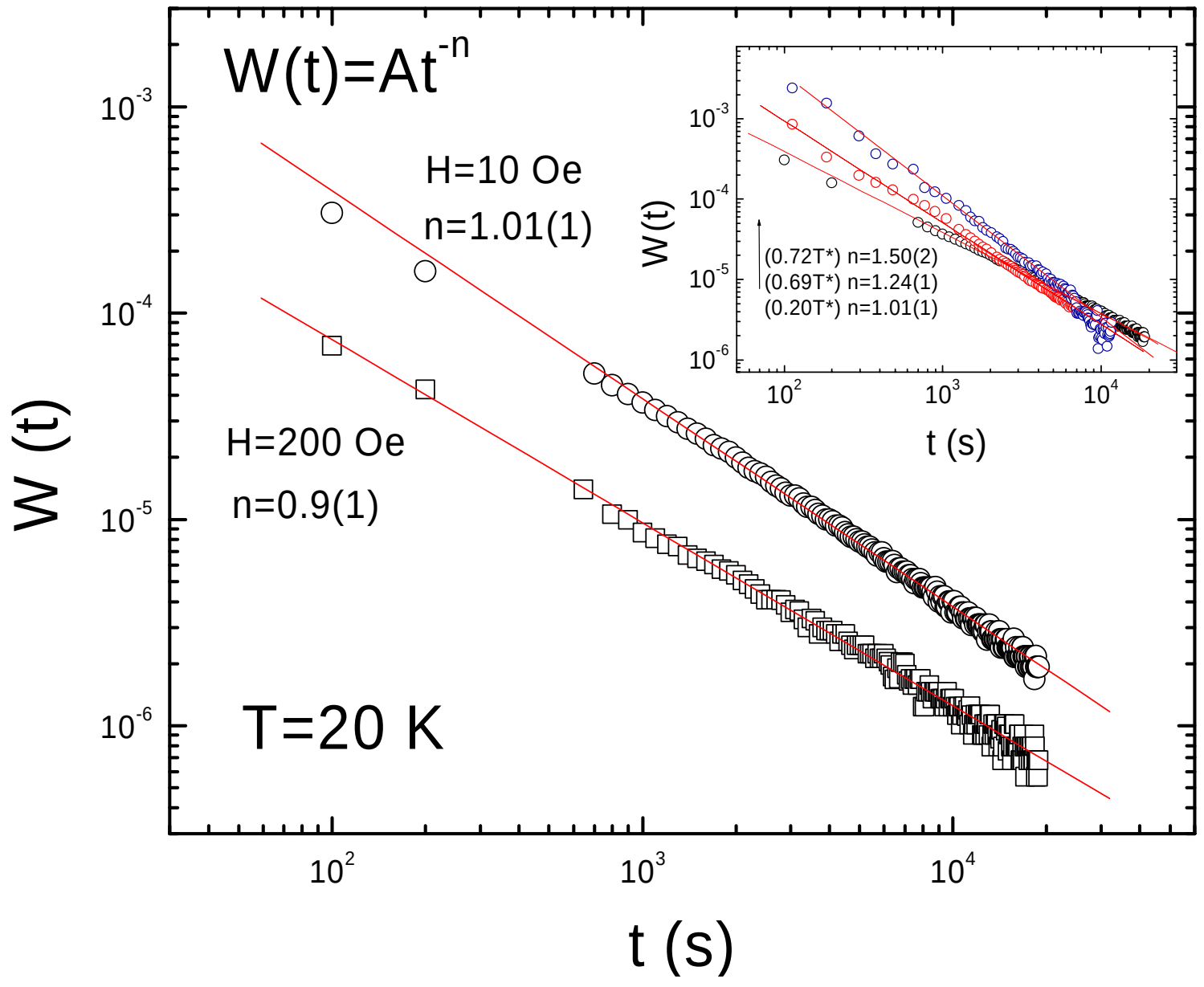

**Figure 4**: Relaxation rate at constant temperature for different cooling fields. The exponent $n \approx 1$ for moderate low fields. *Inset*: The relaxation rate as a function of T. As T* is gradually approached from below *n* increases reflecting an increase in the magnetic interactions between clusters that produce the spin-glass like characteristics.

This is a very important result for the characterization of the PSS, because it discards the competition between DE and SE as the origin of the frustration and the glassy behavior of the PSS (we should expect a constant value of n with T below T* if this were the case). This experiment constitutes a definitive proof to demonstrate that the PSS in manganites must be considered as an assembly of interacting magnetic particles. It is the concentration of particles, which could be controlled by composition and/or magnetic field, which introduces the frustration observed experimentally and hence determines the collective behavior observed in the relaxation of the system.

Another important piece of information comes from the behavior of the nonlinear susceptibility measured from the ac-susceptibility. The negative peak in $\chi_3'$ (Fig. 5) is very broad compared to that of a typical spin-glass. Moreover, the Curie-Weiss dependence of the linear susceptibility and the $T^{-3}$ dependence of the non linear term above T* (Fig. 5, inset) is consistent with the Wohlfarth blocking model for an assembly of interacting magnetic particles [21].

The broadening and displacement of the curve to higher temperatures when a dc field is superimposed to the ac curve [16] is identical to the effect of an increase in the concentration of magnetic particles in a quenched ferrofluid [22].

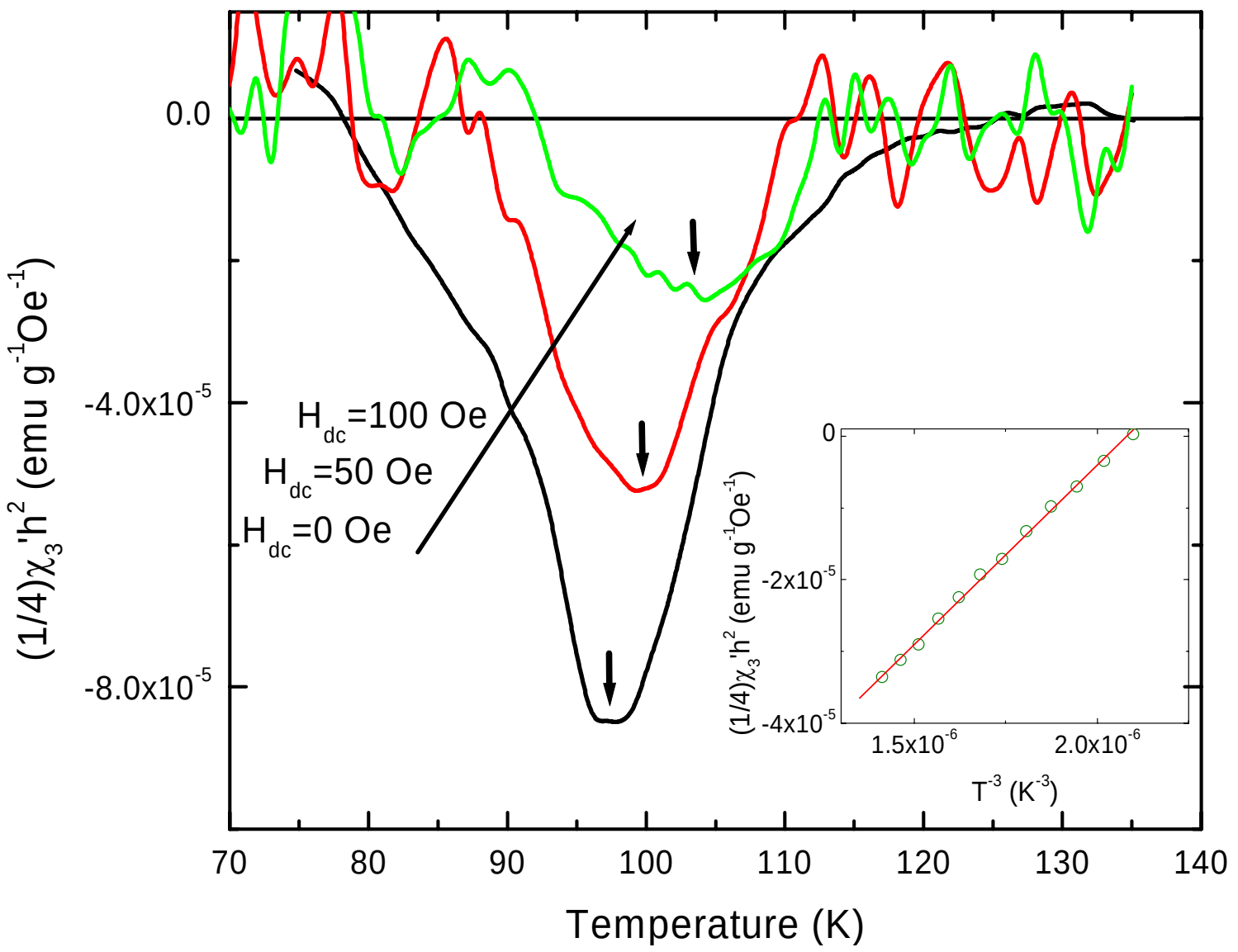

**Figure 5**: The real part of the cubic susceptibility vs. temperature for a field cooled sample (the cooling field is indicated). *Inset*: $T^{-3}$ dependence of the cubic susceptibility above T*.

In summary, we have demonstrated that the PSS observed in manganites close to a metal-to-insulator crossover must be considered as an assembly of interacting magnetic particles, quite similar to a sort of self-generated magnetic colloid, at least from a magnetic point of view. Our experiments demonstrate that it is the interaction between clusters and not the competition between DE and SE what is at the origin of the collective/glassy behavior observed experimentally. The interactions among the clusters can be tuned by composition and/or magnetic field, through the control of the size and concentration of the magnetic clusters. The system, although does not constitute a conventional spin-glass in the sense that it cannot be associated to any of the existing universality classes, shares many of its features below T*. In a simple model, the origin of this magnetic relaxation in the glassy region can be interpreted as a consequence of the strong coupling of the magnetic moments of a dense assembly of clusters.

We would like to stress out that the fact that the T* of the system and the $T_C$ of the individual particles is practically identical, complicates the analysis of the data and a straightforward applicability of the existing models could be inappropriate. We hope our experimental results will attract the interest of future theoretical discussions in this direction. We believe these results are general and should be applicable to other systems close to a first order electronic transition similar to the one described here, like cobaltates, etc.

**Acknowledgments.**

F. R. acknowledges the MC&T of Spain for financial support under the program Ramón y Cajal. This work was financed by FEDER project MAT2001-3749 from the MC&T, Spain.